\documentclass[aps,prl,twocolumn]{revtex4}

\usepackage{amsmath}
\usepackage{amssymb}
\usepackage{amsthm}
\usepackage{psfrag}
\usepackage{graphicx}

\newcommand{\beq}{\begin{equation}}
\newcommand{\eeq}{\end{equation}}
\newcommand{\beqa}{\begin{eqnarray}}
\newcommand{\eeqa}{\end{eqnarray}}

\newcommand{\sm}[1]{{\scriptscriptstyle \rm #1}}

\newcommand{\lp}{\left}
\newcommand{\rp}{\right}

\newcommand{\OO}{{\cal O}}

\newcommand{\eg}{{\em e.g.}}

\newcommand{\DC}{\text{DC}}

\input epsf%
\begin{document}
\title{Holographic Polarons, the Metal-Insulator Transition and Massive Gravity}
\author{Matteo Baggioli\! and Oriol Pujol\`as}

  \affiliation{
 \it Departament de F\'isica and IFAE, Universitat Aut\`onoma de Barcelona,
 Bellaterra 08193 Barcelona Spain}

\begin{abstract}
Massive gravity is holographically dual to `realistic' materials with momentum relaxation. The dual graviton potential encodes the phonon dynamics and it allows for a much broader diversity than considered so far. We construct a simple family of isotropic and homogeneous materials that exhibit an interaction-driven Metal-Insulator transition. The transition is triggered by the formation of polarons  -- phonon-electron quasi-bound states that dominate the conductivities, shifting the spectral weight above a mass gap. We characterize the polaron gap, width and dispersion. \\[-7mm]

\end{abstract}

\maketitle

\thispagestyle{plain}%

{\em Introduction:}
Strongly correlated materials are interesting because they give rise to rich collective behaviour and emergent phenomena. Classic examples  are high-temperature superconductors  and their parent materials, often strange metals and Mott-like insulators \cite{Gebhard,Basov}. The purpose of this Letter is to use the Gauge/Gravity duality (GGD) to shed some light into the collective behaviour 
behind the interaction-driven metal-insulator transition.

The central observation that justifies the use of the GGD beyond the few duality pairs that have been identified so far is that the key features of the AdS/CFT duality \cite{Maldacena:1997re} seem to hold rather generally. 
Conformal Field Theories (CFTs) that admit a gravity dual are believed to be a sub-class of the CFTs with a large N limit and a large `t Hooft-like coupling such that the spectrum of light operators in the CFT reduces to just the stress tensor $T_{\mu\nu}$ and a few more. 
The (quantum) dynamics in these CFTs is expected to simplify enormously into a classical gravitational theory in asymptotically Anti-de Sitter (AdS) space in one more dimension, that plays the role of the renormalization scale. Here we will use this notion of the GGD, working directly in the gravity dual and extracting its CFT interpretation.

An interesting recent development is that in order to model realistic materials that incorporate momentum relaxation, the holographic dual needs to be a massive gravity theory \cite{Vegh:2013sk}. Momentum non-conservation requires to break translation invariance, which is part of a gauged symmetry (diffeomorphism invariance) in the gravity dual. The simplest way to break it is via a graviton mass term (see \cite{Hartnoll:2012rj,Hartnoll:2007ip,Donos:2012js,Horowitz:2012ky} for previous attempts). Importantly, the physical content of this breaking is the introduction of new degrees of freedom. In the present context, these are identified as   the {\em phonons}. \\[-3mm]

{\em Holographic Massive Gravity (HMG):}
There are two ways to define a massive gravity theory. One is a diffeomorphism non-invariant language where we add a potential term for the metric. To fix ideas, let us concentrate on a choice similar to the one suggested in \cite{Vegh:2013sk} 
$$
m^2 V(P_{\mu\nu}g^{\mu\nu})
$$ 
where $P_{\mu\nu} $ is a fixed external matrix that is assumed to project only on the spatial coordinates, $x^i$, probed by the CFT \footnote{For planar materials (2+1 CFTs) the gravity dual coordinates are $x^\mu=\{u, t, x^i\}$, with $u$ dual to the RG scale.}. Thus, $P_{ij}=\delta_{ij}$ and zero otherwise,  which leads to Lorentz non-invariant mass-term. 

This theory propagates extra degrees of freedom. It is very convenient to make them explicit resorting to a diffeomorphism invariant presentation. This was systematized in \cite{Rubakov:2004eb,Dubovsky:2004sg} for general Lorentz non-invariant mass-terms. 
The minimal covariantization in the present case requires a set of scalar fields $\Phi^I$  transforming under an internal Euclidean group of translations and rotations in field-space. The massive gravity action is  recovered as the truncation to  2-derivative operators,
\begin{equation}\label{S}
S_{HMG} = M_P^2\int d^4x \sqrt{-g}
\left[\frac{R}2+\frac{3}{\ell^2}- \, m^2 V(X)\right]
\end{equation}
with 
$X \equiv \frac12 \, g^{\mu\nu} \,\partial_\mu \Phi^I \partial_\nu \Phi^I$ and $\ell$ the AdS radius.
The theory admits solutions where the scalars take on linear vevs 
$\bar \Phi^I = \alpha \,\delta^I_i \,x^i$ with $\delta^I_i$ the Kronecker delta.
On these solutions, the previous projector $P_{\mu\nu}$ is identified as $\propto\partial_\mu\bar \Phi^I\partial_\nu\bar \Phi^I$. This procedure was also described in \cite{Andrade:2013gsa,Taylor:2014tka} for the potential assumed in \cite{Vegh:2013sk}.

Our first point is that Ref. \cite{Vegh:2013sk} and most of the literature on HMG to date, unnecessarily restrict to a very narrow family of potentials. They assume formally the same potential for the metric as in the dRGT massive gravity \cite{deRham:2010kj}. That choice has a number of advantages when the external metric $P_{\mu\nu}$ is Minkowskian, but in the Lorentz non-invariant case that is relevant for condensed matter, the set of consistent choices is much broader \cite{Rubakov:2004eb,Dubovsky:2004sg}.

Let us discuss the constraints on the form of $V(X)$. First, we require absence of ghosts and other pathological instabilities. Around configurations with rotational symmetries such as the black branes \eqref{backg}, it is convenient to separate the analysis of  linear perturbation in scalar, vector and tensor modes.  The full analysis of the scalar sector is a bit lengthy because of the profusion of scalar modes (in the metric, vector and Goldstones). Instead, we will perform now the analysis  in the `decoupling limit' where we turn on only Goldstone fluctuations, which should be
valid for small $m$.
We perturb the Goldstones, $\Phi^I= \bar\Phi^I + \phi^I$, and expand the Lagangian to second order. Dropping tadpole terms and the distinction between $I$ and $i$ indices, one gets
\beq\label{L2}
  V'(\bar X) \partial_\mu \phi^i \partial^\mu \phi^i + \bar X V''(\bar X) (\partial_i\phi^i)^2 
\eeq
Absence of ghosts, then,  leads to monotonic potentials
\beq\label{C1}
V'(\bar X)>0~.
\eeq 
The local (sound) speed of longitudinal phonons
is 
\beq\label{C2} 
c_S^2 =1+\frac{\bar X V''(\bar X)}{V'(\bar X)} >0~.
\eeq
While this is not yet the speed of any physical excitation, it is safest to require it to be everywhere positive to guarantee the absence of gradient instabilities \footnote{The form of $V(X)$ that sets $c_S^2=0$ is $V(X)\propto \log X$.}.

No further constraints arise from the vector and tensor sectors nor at nonlinear level in the scalars, which is not surprising since  a substantial advantage of the Lorentz non-invariant mass terms is that they can be free from the Boulware-Deser ghost \cite{Rubakov:2004eb,Dubovsky:2004sg}. 
We obtain further constraints by requiring that the theory admits asymptotically AdS solutions with $X\to0$, implying that $V(0)=0$. In addition, requiring that the Goldstone sector is weakly coupled for $X=0$ requires that $V'(0)$ is not too small.

All in all, an economic and safe way to satisfy all constraints is to assume that 
$V(X)$ is monotonous and linear close to $X=0$. To fix ideas, we shall consider potentials of the form $V_N(X)=X+{\beta } X^N/N$ with $\beta>0$ and $N>1$.

In any asymptotically AdS solution (like the planar black branes in Eq.~\eqref{backg}), $\bar X =\alpha^2 u^2$ and close to the AdS boundary ($u=0$) the Goldstone gradient $\bar X$ vanishes. This implies that the mass term for  metric modes is also vanishing. So this is a weaker form of massive gravity than is usually discussed in cosmology -- the Compton wavelength is at most of order the curvature radius of the spacetime. 
In the CFT interpretation, the stress tensor $T_{\mu\nu}$ does not develop an anomalous dimension \footnote{Generating an anomalous dimension requires $V\sim \log X$, which upsets longitudinal phonons and AdS asyptotics.}.  
Still, as shown in \cite{Vegh:2013sk} this is enough to lead to momentum relaxation in the CFT. 
The good news is that the Goldstone sector required for that is very healthy -- again, because it is enough that the mass term is `active' only in the interior of AdS. This is why the form of the potential (the kinetic function of the Goldstones) is almost unconstrained. For the same reason, one can be confident that a proper analysis of the scalar sector will not change much the conditions \eqref{C1}, \eqref{C2}.\\[-3mm]

{\em Impurities and phonon dynamics:}
Let us now translate into CFT language. Demanding that the CFT includes momentum relaxation we ended up with a rather general form of massive gravity. This theory contains a universal sector consisting of the Goldstone degrees of freedom. In CFT language, these are a multiplet of operators $\OO^I$ with internal shift symmetries and which are somehow related with phonons and impurities \cite{Vegh:2013sk}. A consistent interpretation seems to be that the strength $\alpha$ of the linear vevs $\bar\OO^I = \alpha \delta^I_i x^i$ is the density of homogeneously-distributed impurities. The fluctuations $\delta\OO^i$ around this distribution are CFT operators that create phonon excitations.  

Let us now look at the quadratic action \eqref{L2} more closely.
Turning to the canonically normalized field $\sqrt{V'(\bar X)} \phi^i$, one sees that both transverse and longitudinal phonons develop a `scale-dependent' mass ($\bar X\sim u^2$) 
\beq\label{phimass}
M_\Phi^2(u)=(\partial \bar a)^2+\Box {\bar a} = {f(u\partial_u \bar a)^2\over \ell^2}+{u^2\partial_u(f \partial_u \bar a)\over \ell^2}
\eeq
where $\bar a(u) = {1\over 2} \log V'(\bar X)$ 
and in the second equation we specialized to the BB metric \eqref{backg}.
We note a few interesting properties: i) $M_\Phi \neq0$ only for non-linear $V$; ii) close to the UV boundary, $M_\Phi^2 \sim {V''(0)\over V'(0)} u^2$ so $\delta O^I$ admits a unitary standard quantization with zero anomalous dimension; and iii) on the horizon ($f=0$), 
$$
M_\Phi^2 \sim - {|f'(u)|} {V''(\bar X) \over V'(\bar X)} \;\Big|_{\rm horizon} \;< 0~. 
$$ 
It follows that a large enough  ${V'' / V'}$ leads to an infrared instability near the horizon. This triggers a massive shift of the spectral weight in the response functions, with the  associated formation of a gap and eventually a sharp resonance \footnote{This mechanism can be seen to take place within the validity of the effective field theory, see \cite{future}.}. Physically, this represents a self-supported lattice distortion. In the presence of charge carriers, 
such phonon resonant modes trap the carriers, forming proper polarons, and can effecively suppres the DC conductivity. (See \cite{Devreese-Alexandrov,Gebhard,Basov} for reviews on the polarons concept and its applications.)
The above instability, then, is the basic mechanism of {\em polaron formation} in its holographic incarnation.  
The numerical computation of the  optical conductivity confirms this picture, see Fig.~1. 
Indeed, for sizeable $V'' / V'$ we find a spectral weight transfer into a `mid-infrared' (polaron)  peak.

The instability $M_\Phi^2<0$ close to the horizon is reminiscent of holographic superconductivity \cite{Hartnoll:2008vx}. But there are notorious differences. Here, the instability does not lead to any condensation -- the order parameter is not the expectation value of an operator. 
Instead, one would identify the order parameter as the energy gap $\omega_0$, the polaron rest-energy once it is formed. Hence, the transition proceeds by a re-arrangement of the degrees of freedom. 

Importantly, polaron formation is perfectly compatible with absence of gradient instability because $V''/V'$ needs to be positive.  Notice that in this case \eqref{C2} is superluminal. However, this does not imply that there are superluminal excitations. The polaron quasiparticles below are localized close to the horizon. Their physical speed, then, receives an additional redshift factor that can render them subluminal. It seems quite generic, though, that whenever polarons form, the longitudinal phonons/polarons should be faster than transverse ones.\\[-3mm]

{\em Holographic Metal-Insulator transition (MIT):}
In order to model a transition between itinerant and polaron-localized excitations, we need to add one ingredient, the charge carriers. So we assume that the CFT also contains a conserved current operator $J_\mu$. This is implemented in the gravity dual by adding to the model a Maxwell field,
\begin{equation}
S = S_{HMG} -\frac{M_P^2}{4} \int d^4x \sqrt{-g} F_{\mu\nu}F^{\mu\nu}
\end{equation}
with $F_{\mu\nu}=\partial_{[\mu}A_{\nu]}$. As usual, the subleading mode of $A_\mu$ towards the AdS boundary is the dual $J_\mu$.
To introduce a finite density of charge carriers $\rho$, we turn on a chemical potential $\mu$. The dual of the CFT ground state at finite $\mu$, temperature $T$ and impurity strength $\alpha$ is a planar black brane (BB) with $A_t(u\to0)=\mu$, $\Phi^I= \alpha \,\delta^I_i\, x^i$ and 
\beq\label{backg}
ds^2=\frac{\ell^2}{u^2} \left[\frac{du^2}{f(u)} -f(u)dt^2 + dx^2+dy^2\right] ~.
\eeq
Here, $u$ is the holographic coordinate dual to the renormalization scale $\mu\sim 1/u$. 

\begin{figure}[t]
	 \includegraphics[width=7.5cm]{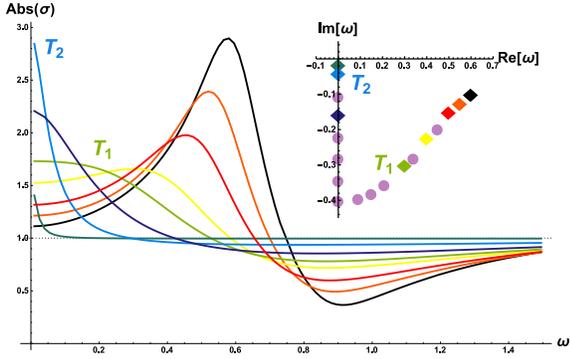}\\[-3mm]
  \caption{{\footnotesize Development of polaron formation as temperature is decreased.  $T=1,\,0.46 (T_2),\, 0.35,\,0.3 (T_1),\,0.27,\,0.22,\,0.17,\,0.04$. 
  Inset: motion of the lightest quasi-normal-mode (QNM) for the same temperatures (corresponding colours) and some intermediate additional values (purple dots). At large $T$, the QNM separates from the real axis with decreasing $T$, until it collides with the next QNM (near $T_1$) and forms a pair of conjugated poles with positive and negative real parts -- the polaron particle/anti-particle poles. (Similar QNM collisions have been observed in \cite{Davison:2014lua}.)}}
   ~\\[-8mm]
\end{figure}

The Maxwell equations set $A_t =\mu - \rho \,u$  with $\rho$ the charge-carrier density. The Einstein equations reduce to 
\beq
u f'(u)= -3 +3 f +\rho^2 u^4 /(2\ell^2)  + (m \ell)^2\,V\left({\alpha^2 u^2 / \ell^2}\right)
\eeq
The solution for general $V$ is
\begin{equation}
f(u)= u^3 \int_u^{u_H} dv\;\left[ \frac{3}{v^4} -\frac{\rho^2}{2\ell^2}-\frac{(m\ell)^2}{v^4}\, 
V\left({\alpha^2 v^2\over \ell^2}\right) \right]
\end{equation}
where $u_H$ stands for the location of the BB horizon.  Regularity of the gauge field on the horizon requires  $\rho = {\mu \over u_H}$, and that the temperature is (from now on, $\ell=1$)
\begin{equation}
T=-\frac{f'(u_H)}{4\pi}=\frac{6 - {\mu^2 u_H^2} -  2m^2 V\left(\alpha^2 u_H^2 \right) }{8 \pi u_H}~.
\end{equation}
The energy density is given by $f'''(0)/6$ which can be expressed in terms of $\mu$, $\alpha$ and $u_H$. In practice, though, it is more convenient to trade the energy density in favor of $u_H$, which relates  to the entropy density ($s\propto u_H^{-2}$). \\[-3mm]

\begin{figure}[t]
 	\[
	\begin{array}{ccc}
	    \includegraphics[width=4cm]{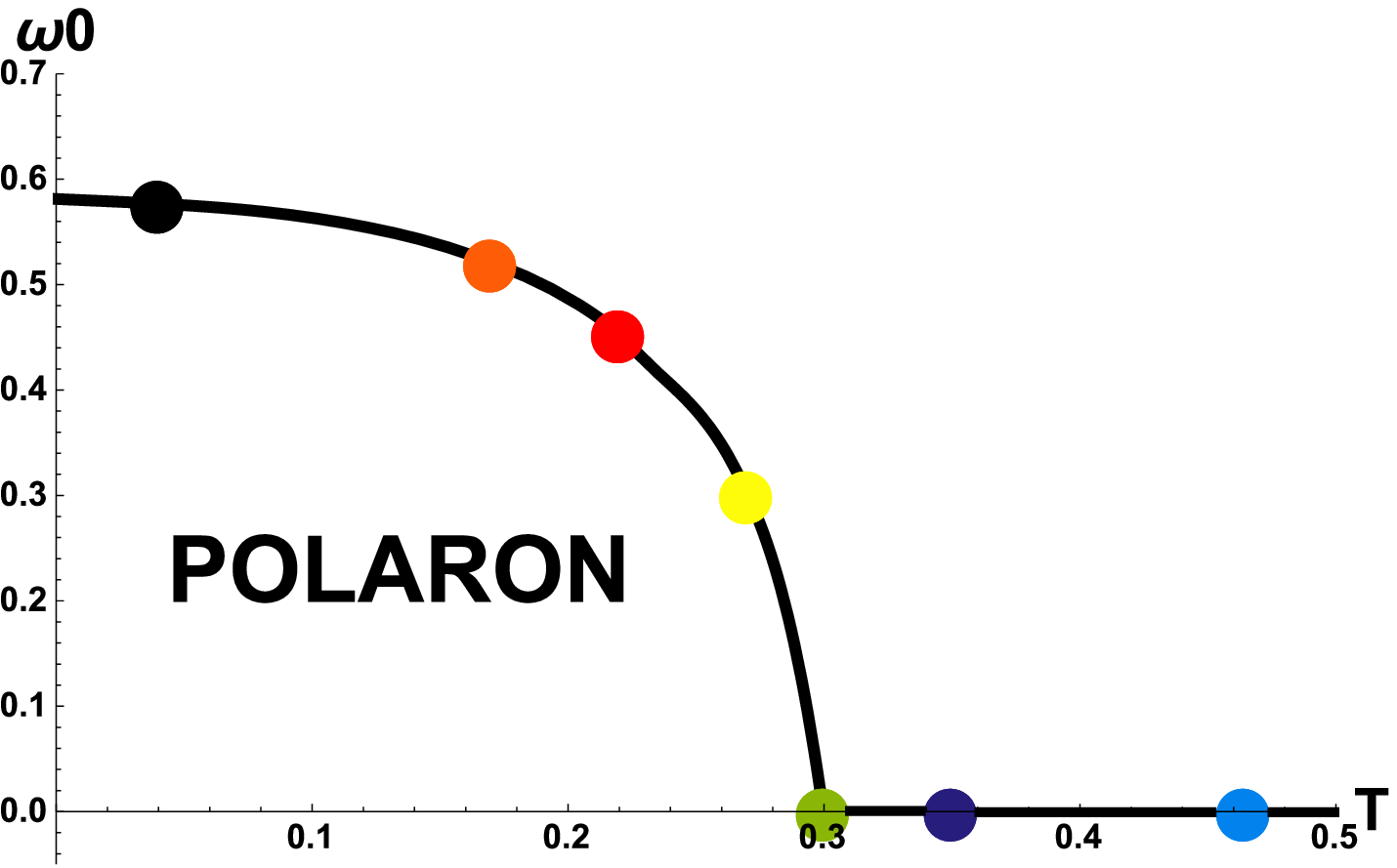}~~~~
   &   \includegraphics[width=4cm]{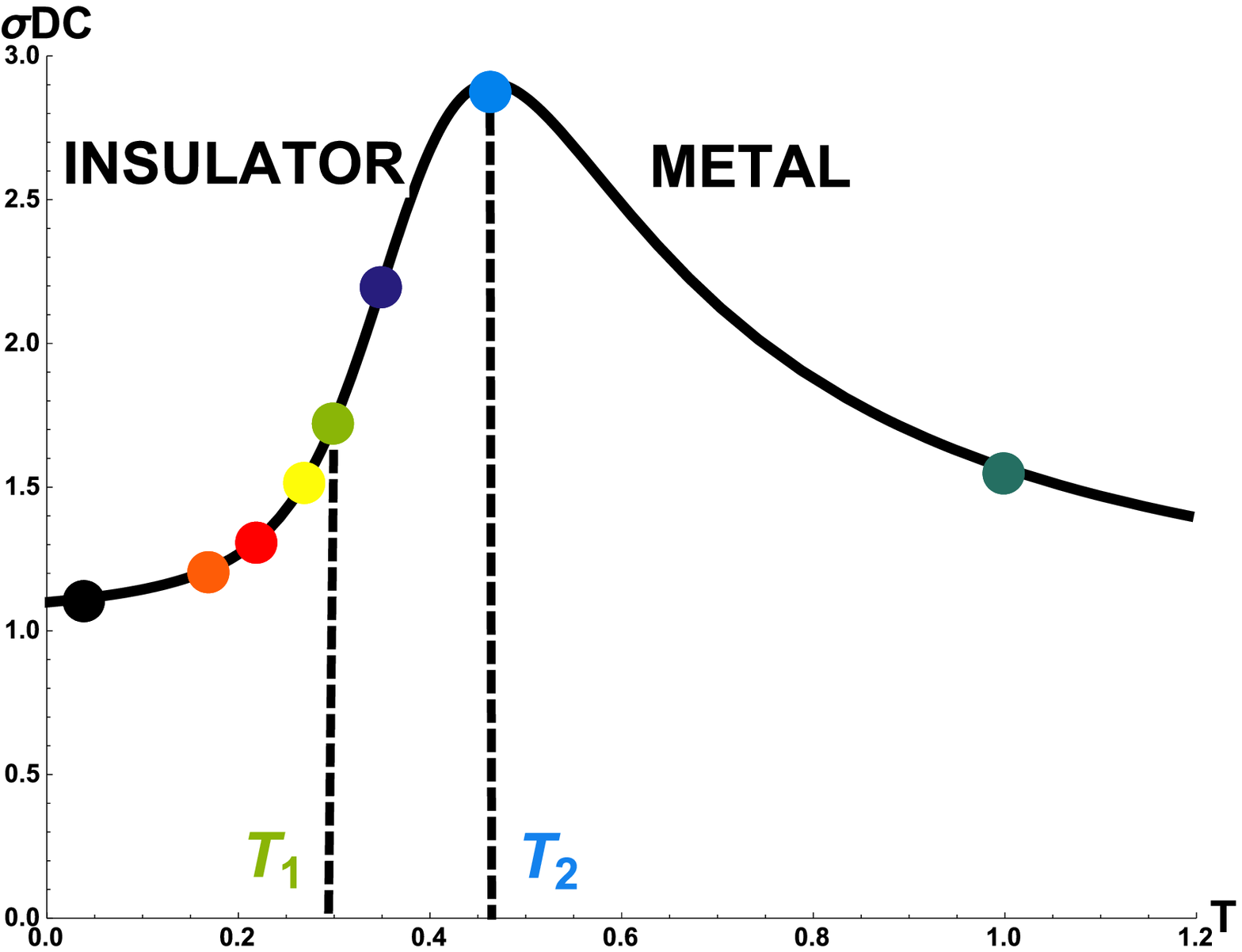}\\
   \sm{(a)}&\sm{(b)}\\[-6mm]
	\end{array}
	\]
  \caption{{\footnotesize (a) The polaron energy-gap $\omega_0$ as a function of $T$, extracted numerically as the peak-position in ${\rm Abs}[\sigma(\omega)]$. $\omega_0$ detaches from 0 at $T_1$. (b) DC conductivity $\sigma_{\DC}(T)$ as a function of $T$. The continuous line is the analytic result \eqref{DC} and the dots are the numerical computation. One observes two critical temperatures: the  the maximum in $\sigma_{\DC}(T)$ marking the metal/insulator transition ($T_2$), and the  polaron formation $T_1$. We find $T_1<T_2$ generically. }}\label{fig:GL}
~\\[-5mm]
\end{figure}

{\em Optical conductivities:}
Let us now discuss the behaviour of small excitations around the BBs above. We  perturb the previous solutions by setting $A_\mu=\bar A_\mu+a_\mu$, $g_{\mu\nu}=\bar g_{\mu\nu}+h_{\mu\nu}$ and $\Phi^I=\bar \Phi^I+\phi^I$ with bars denoting the background solutions, and linearize. In the remainder, we concentrate on transverse vector modes (\eg\, $\partial^i a_i =0$ \footnote{In the homogeneous limit, the transverse vector variables become just constant vectors under rotations.}).
Vector perturbations are encoded in
$ a_i, \, \phi_i, \, h_{ti}, \, h_{ui} \; {\rm and} \;  h_{ij}\equiv \frac{1}{u^2}\partial_{(i} b_{j)}$. 
Aside from $a_i$, we use the gauge-invariant combinations 
\beq
T_i \equiv u^2 h_{ti} - {\partial_t \phi_{i} \over \alpha} \; , \;
U_{i}  \equiv  f(u)\big[h_{ui} -  {\partial_u \phi_{i}\over \alpha u^2}\big] \; , \;
B_i \equiv   b_i -{\phi_i\over\alpha} ~.\nonumber
\eeq
The linearized equations  can be written as \footnote{$T_i$ is constrained by an algebraic equation, see \cite{future}.}
\begin{eqnarray}
{\partial_u}\lp( f \, \partial_u \;a_i \rp)+\Big[\frac{\omega^2}{f} -k^2-2 u^2 \rho^2\Big] a_i =~~~~~~~~~~~~~~~~~~~~&&\cr
 \frac{i \rho \,u^2 (2\bar m^2+k^2)  }{\omega} U_i \,
-\frac{i f \rho\, k^2 }{\omega} \partial_u B_i  \, , && \nonumber \\[0mm]
\frac{1}{u^{2}}\partial_u\lp[ \frac{f u^2}{\bar m^2}  \partial_u \big(\bar m^2 U_i\big)\rp]
+\Big[\frac{\omega^2}{f} -k^2-2 \bar  m^2\Big] U_i  = ~~~~~~~~&& \cr
 - 2 i \,\rho \,\omega \,a_i + \frac{f' k^2}{u^2} B_i   \,, && \nonumber \\[0mm] 
k\Big\{u^2{\partial_u}\Big(\frac{f}{u^2} \, \partial_u \,B_i \Big)
+\Big[\frac{\omega^2}{f} -k^2-2 \bar m^2 \Big] B_i 
=-2 \frac{\bar m'}{\bar  m} U_i  \Big\}\,,&&\nonumber
\end{eqnarray}
where we introduced $\bar m^2(u) = \alpha^2 m^2 V'(\alpha^2 u^2)$.

The optical conductivities  can be extracted from the numerical integration of these equations. One imposes infalling boundary conditions (dual to the retarded Green function prescription). For the electric conductivity, one imposes that $U_i=0$ at $u=0$ and reads off the optical conductivity as 
$\sigma(\omega) = \frac{\partial_u a_j}{i\,\omega\,a_j}\big|_{u\to0}$ (no summation over repeated indices). 

The results are plotted in the Figures for $V(X)=X+X^5$, $(m\ell)^2=0.05$, $\rho=1$ and  $\alpha=\sqrt2$. Polaron formation is seen in Fig.~1 as $T$ decreases \footnote{At $T=0$ there is a polaron formation quantum phase transition driven by increasing $V''(\bar X)$ at the horizon.}. Fig.~2 shows the polaron gap (order parameter), $\omega_0(T)$, and $\sigma_{\DC}(T)$. Fig.~3 shows the polaron dispersion relation, fitting well to $\omega=\omega_0+k^2/(2m_*)$ at low $k$.

In the homogeneous limit $k\to0$ the gauge-invariant variable $B_i$ decouples and it is consistent to set $B_i=0$. Following \cite{Blake:2013bqa} one arrives at a formula for $\sigma_{\DC}$ \cite{Davison:2013jba,Blake:2013bqa},
\begin{equation}\label{DC}
\sigma_{\text{DC}} = 1+ \frac{\rho^2 u_H^2}{\alpha^2 m^2 V'(\alpha^2 u^2_H)}~.
\end{equation}
The plot $\sigma_{\text{DC}}(T)$ in Fig 2 (b) clearly displays a change from metallic ($d\sigma/dT<0$) to insulating ($d\sigma/dT>0$) behaviour. Let us emphasize that we take $\rho$ constant in order to isolate the dynamics that drives the transition\footnote{Keeping $\mu$ constant, a MIT can also be obtained  choosing, e.g., $V(X)=X^\epsilon+\beta X^N$ with $0<\epsilon<1$, $N>2$.}. 
 From \eqref{DC}, one can extract analytic expressions for the critical temperature, $T_2$, where $d\sigma/dT=0$. The condition becomes $\bar X V''(\bar X)= V'(\bar X)$, 
in agreement with the analysis above in terms of $M_\Phi^2<0$ near the horizon. For the benchmark model $V_N(X)$ above, one finds  $\alpha \,u(T_2)= [(N-2)\beta ]^{-{1\over 2(N-1)}}$, showing that one needs a high enough exponent, $N>2$. Models with $N<2$, tend to give incoherent metals with no MIT.

\begin{figure}[t]
	 \includegraphics[width=7.0cm]{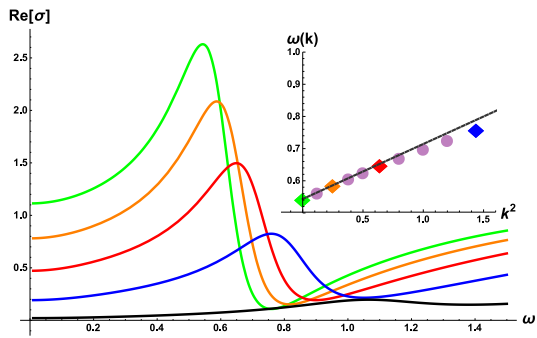}\\[-3mm]
  \caption{{\footnotesize 
  Motion of the polaron peak  with wavenumber. $T=0.04$, $k=0,\,0.5,\,0.8,1.2,\,2$. Inset: the extracted dispersion relation.
   }}\label{BH}
   ~\\[-8mm]
\end{figure}

These results indicate that one can distinguish 2 different critical temperatures  (numerically not far from each other): $T_2$ marks the maximum in $\sigma_{\DC}$ and $T_1$ marks the polaron formation (defined as when the peak position, $\omega_0$, in  ${\rm Abs}(\sigma(\omega))$ separates from $0$). Fig.~2(a) suggests that the phase transition across $T_1$ is of second order, even if there is no symmetry breaking. One expects that the spatial correlation length becomes finite only below $T_1$, but this requires a proper analysis. 

We must note the quite intriguing intermediate range $T_1<T<T_2$, showing insulating behaviour yet without polarons. A possible interpretation is that polaron form first incoherently for $T_1<T<T_2$ and coherently for $T<T_1$. See \cite{Mannella-Zaanen,Merino-McKenzie,Fratini-Ciuchi} for similar ideas. \\[-3mm]

{\em Discussion:}
In summary, we have seen that a simple nonlinear extension of holographic massive gravity captures interesting features of correlated materials such as  polaron-localization and a (phonon-)interaction-driven Metal-Insulator transition (MIT). The spectrum of excitations around simple black brane solutions include clear polaron-like quasi-localized states close to the MIT transition. Unlike previous holographic models \cite{Donos:2012js,Donos:2013eha,Ling:2014saa,Donos:2014oha}, our MIT takes place in homogeneous and isotropic materials and it does not rely strongly on the running of the electromagnetic coupling (as in \cite{Donos:2013eha,Gouteraux:2014hca}).
See also \cite{Amoretti,Davison:2014lua,Liu:2012tr} for the connection between holographic models and quasi-particle descriptions.
Several other aspects of the transition are going to be presented elsewhere \cite{future}, but one can advance potential correlations: when polarons form, the piezo-electric response is large, and longitudinal modes should be faster than transverse ones. 

Let us emphasize that polarons are thought to play an important role in cuprate superconductors and colossal magnetoresistance in manganites, see \eg \,\cite{Devreese-Alexandrov,Basov,Mannella-Zaanen,Palstra,Mannella,Massee}. Additionally, there is a considerable body of experimental evidence that polarons do occur in a variety of materials \cite{Devreese-Alexandrov,Palstra,Mannella,Ronnow,Billinge,Massee,Koschorreck,Kohstall}. 
It would be interesting to understand how much can be learned  from the new holographic handle on polaron physics. From our perspective, there seems to be potentially many other massive gravity phases that could be similarly relevant for this and and perhaps other condensed matter applications. We hope to return to these questions soon.

\paragraph*{Acknowledgements}
We thank A. Pomarol, N. Magnoli and S. Sibiryakov for useful discussions. We acknowledge support from MINECO under grant FPA2011-25948,  DURSI under grant 2014SGR1450 and Centro de Excelencia Severo Ochoa program, grant SEV-2012-0234. OP is supported by a Ramon y Cajal fellowship (MICINN-RyC program).

\end{document}